%% file: main.tex
\documentclass[fleqn,10pt]{wlscirep}
\usepackage[T1]{fontenc}
\usepackage{subfig}
\usepackage{bbold}
\usepackage{amsmath}
\usepackage{amsfonts}
\usepackage{bm}
\usepackage{mathrsfs}
\usepackage{epstopdf}
\usepackage{hyperref}
\newcommand{\scs}{\scriptscriptstyle}

\title{Fano feature induced by a bound state in the continuum via resonant state expansion}

\author[1,2,+,*]{Pavel S. Pankin}
\author[1,2,+]{Dmitrii N. Maksimov}
\author[3]{Kuo-Ping Chen}
\author[1,2]{Ivan V. Timofeev}

\affil[1]{Kirensky Institute of Physics, Federal Research Center KSC SB
RAS, 660036, Krasnoyarsk, Russia}
\affil[2]{Siberian Federal University, 660041,  Krasnoyarsk, Russia}
\affil[3]{Institute of Imaging and Biomedical Photonics, National Chiao Tung University, 71150, Tainan, Taiwan, ROC}
\affil[*]{pavel-s-pankin@iph.krasn.ru}
\affil[+]{these authors contributed equally to this work}

\begin{abstract}
We consider light scattering by
an anisotropic defect layer embedded into anisotropic photonic crystal in the spectral vicinity of an optical bound state in the
continuum (BIC). Using a resonant state expansion method we derive an analytic solution for reflection and transmission amplitudes.
The analytic solution is constructed via a perturbative approach with the BIC as the zeroth order approximation.
The solution is found to describe the collapsing Fano feature in the spectral vicinity of the BIC.
The findings are confirmed via comparison against direct numerical simulations with the Berreman transfer matrix method.
\end{abstract}
\begin{document}

\flushbottom
\maketitle
%
%
\thispagestyle{empty}

\section{Introduction}

Theoretical insight into resonant response from optical systems,
including photonic-crystalline resonators \cite{Joannopoulos11}
and resonant metasurfaces \cite{Yang15}, is of big importance in
photonics \cite{Miroshnichenko10, Lalanne18}. Very unfortunately
only a few systems generally allow for a tractable analytic
solution providing intuitively clear and mathematically exact
picture, such as, e.g., the celebrated Mie-Lorenz theory
\cite{Stratton41}. Thus, in the field of optics the resonant
scattering quite often can only be understood in terms of the
temporal coupled mode theory (TCMT) \cite{Haus, Fan03, Suh04}. The
TCMT is a phenomenological approach that maps the scattering
problem onto a system of field driven lossy oscillators.
Mathematically, the problem is cast in the form of a system of
linear differential equations. The coefficients of the system
account for both "internal" modes of the resonant structure as
well as for the coupling of the "internal" modes to incoming and
outgoing waves. The interaction with the impinging light is
understood in terms of "coupled modes" which are populated when
the system is illuminated from the far-zone. The elegance of the
TCMT is in its simplicity and the relative ease in establishing
important relationships between the phenomenological coefficients
solely from the system's symmetries and conservation laws
\cite{Fan03, Suh04, Ruan09, Ruan12}. However, despite its numerous
and successful applications, the TCMT generally
 relies on a set of fitting parameters.
 Moreover, the mathematical foundations of the TCMT remain vague
since the theory neither gives an exact definition of the "coupled mode", nor a clear recipe for
such a "coupled mode" to be computed numerically.

Historically, the problem of coupling between the system's eigenmodes to the scattering channels
with the continuous spectrum has attracted a big deal of attention in the field of quantum mechanics
\cite{physrep, Dittes,  Ingrid}. One of the central ideas was
the use of the Feshbach projection method \cite{Dittes,Chruscinski13} for mapping the problem onto the Hilbert
space spanned by the eigenstates of the scattering domain isolated from the environment. Such approaches have met with a limited
success in application to various wave related set-ups, including quantum billiards \cite{Stockmann,Pichugin, Stockmann1},  tight-binding models \cite{SR},
potential scattering \cite{Savin}, acoustic resonators \cite{Maksimov15},
nanowire hetrostructures
\cite{Racec09}
and, quite recently, dielectric resonators \cite{Gongora17}. Besides its mathematical complexity there are
two major problems with the Feshbach projection method: First, the eigenmodes of the isolated systems are in general
not known analytically; therefore, some numerical solver has most often to be applied. Furthermore, the computations of such
eigenmodes requires some sort of artificial boundary condition
on the interface between the scattering domain and the outer space. Quite remarkably the convergence
of the method is shown to be strongly affected by the choice of the boundary condition on the interface \cite{Pichugin, Lee, Schanz}.

In the recent decades we have witnessed the rise of efficient
numerical solvers utilizing perfectly matched layer (PML)
absorbing boundary conditions \cite{Berenger94, Chew94}. The
application of perfectly matched layer has rendered numerical modelling of wave
propagation in open optical, quantum, and acoustic systems
noticeably less difficult allowing for direct full-wave
simulations even in three spatial dimensions. On the other hand,
the application of PML also made it possible to compute the
eigenmodes and eigenfrequencies of wave equations with
refletionless boundary conditions. Such eigenmodes come under many
names including  quasinormal modes \cite{Lalanne18}, Gamow states
\cite{Civitarese04}, decaying states \cite{More71}, leaky modes
\cite{snyder2012optical}, and resonant states \cite{Muljarov10}.
The availability of the resonant states has naturally invited
applications to solving Maxwell's equations via series expansions
giving rise to a variety of resonant state expansion (RSE) methods
\cite{Lalanne18}. One problem with the resonant states is that
they are not orthogonal in the usual sense of convolution between
two mode shapes with integration over the whole space
\cite{Ingrid, Kristensen13}. This can be seen as a consequence of
exponential divergence with the distance from the scattering
center \cite{Lalanne18}. Fortunately, both of the normalization
and orthogonality issues have recently been by large resolved with
different approaches, most notably through the PML
\cite{Sauvan13}, and the flux-volume (FV) \cite{More71,
Muljarov10} normalization conditions.

In this paper we propose a RSE approach to the problem of light scattering by
an anisotropic defect layer (ADL) embedded into anisotropic photonic crystal (PhC) in the spectral vicinity of an optical BIC.
Although BICs are ubiquitous in various optical systems  \cite{Hsu16, Koshelev19}, the
system under scrutiny is the only one allowing for an exact full-wave analytic solution for an optical BIC \cite{Timofeev18}.
By matching the general solution of Maxwell's equation within the ADL to both evanescent and propagating solutions in the PhC
\cite{Rytov1956,
YarivYeh1984bk, ShiTsai1984_PBG, CamleyMills1984_PBG} we find the eigenfield and eigenfrequency of the resonant mode family limiting to the BIC under variation
of a control parameter.
Next, for finding the scattering spectra we apply the spectral representation of Green's function in terms the FV-normalized resonant states \cite{Muljarov10}.
This is a well developed approach which has already been applied
to both two\cite{Doost13}- and  three\cite{Doost14}-dimensional optical systems. The approach has also been recently extended to magnetic, chiral,
and bi-anisotropic optical materials \cite{Muljarov18} as well as potential scattering in quantum mechanics \cite{Tanimu18}. Remarkably, so far RSE methods
have been mostly seen as a numerical tool. Here we show how
a  perturbative analytic solution can be constructed in a closed form within the RSE framework. Such a perturbative solution uses the BIC as the zeroth order approximation and,
very importantly, is capable of describing the collapsing Fano resonance
\cite{Kim, Shipman, SBR, Blanchard16, Bulgakov18a, Bogdanov19} in the spectral vicinity of the BIC.
We mention in passing that due to the fine control of Fano line-shapes
optical BICs has been recently viewed as an efficient
instrument in design of narrowband optical filters \cite{Foley14,Cui16,Doskolovich19,Nguyen19}. We shall see that the
analytic solution matches the exact numerical result to a good accuracy.

\section{The System}
\begin{figure*}
\begin{minipage}[b]{0.49\textwidth}
\center{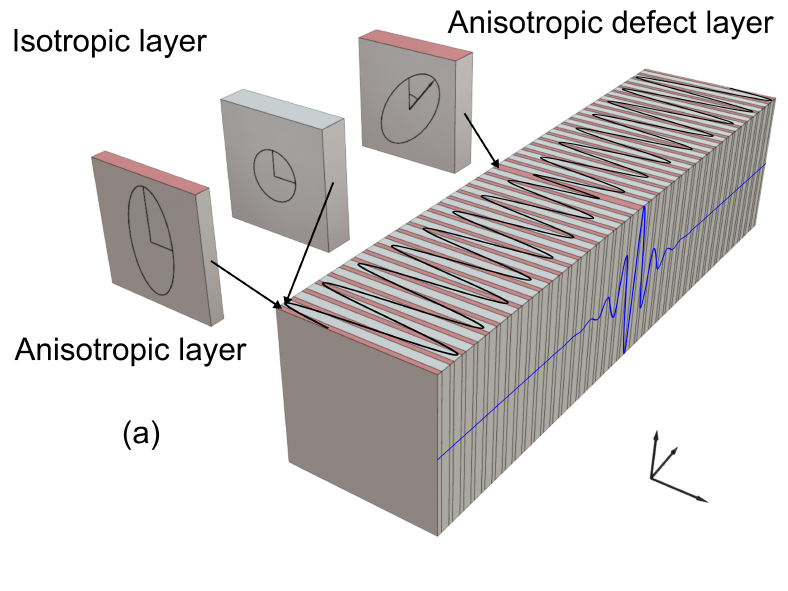}
\end{minipage}
\begin{minipage}[b]{0.49\textwidth}
\centering\includegraphics[scale=1]{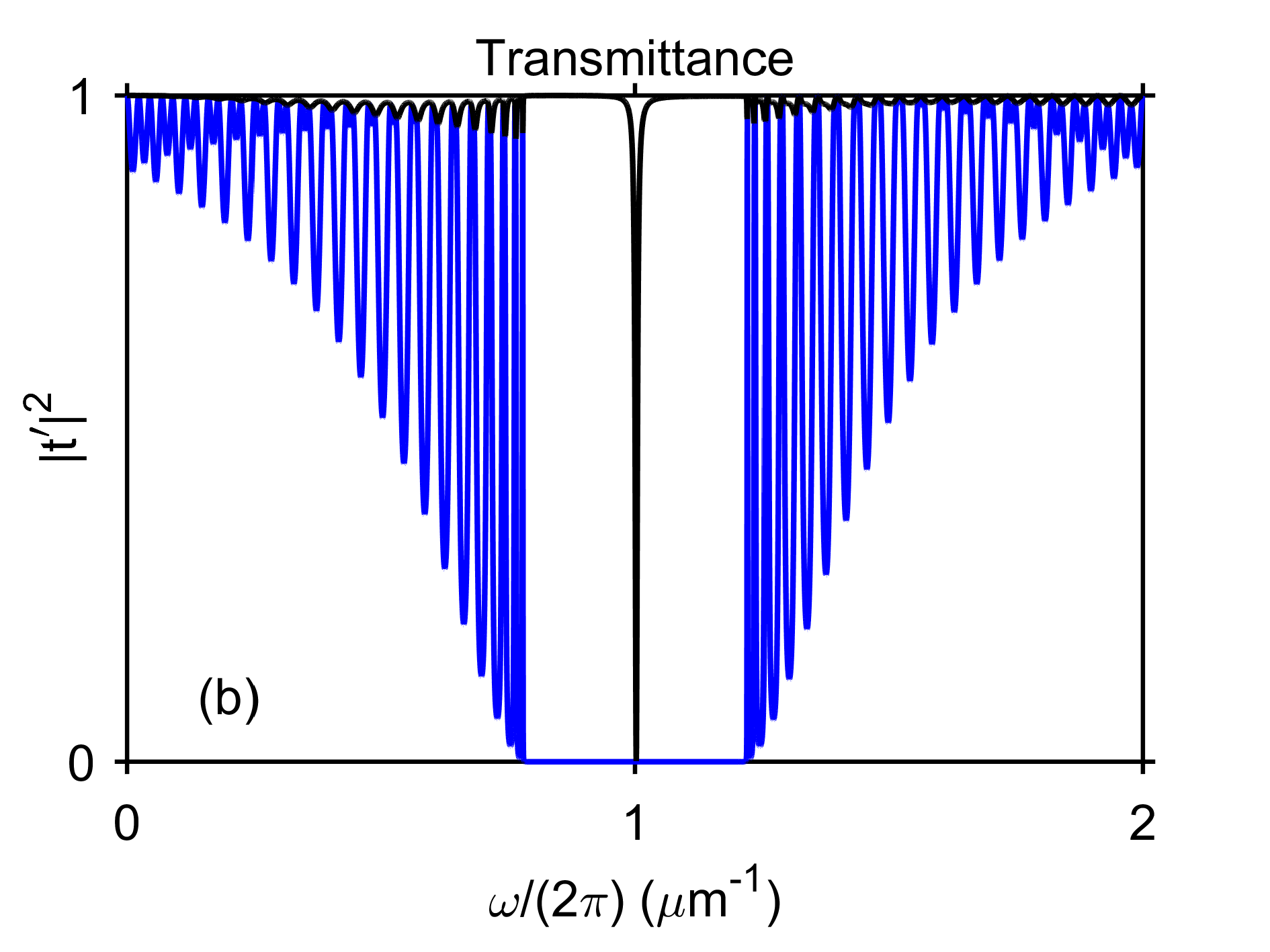}
\end{minipage}
\caption{
{(a) One-dimensional PhC structure stacked of alternating layers
of an isotropic dielectric material with permittivity $\epsilon_o$
(gray) and an anisotropic material with the permittivity
components $\epsilon_o$ and $\epsilon_e$ (pink). An anisotropic
defect layer with a tuneable permittivity tensor is inserted in
the center of the structure. The analytic solution for the
quasi-BIC mode profile 
is plotted on top and right sides of the stack: the $x$-wave component
Re$(E_x)$ -- blue, the $y$-wave component Re$(E_y)$ -- black. (b)
The transmittance spectra $|t^{\prime}|^2$ of $x$- (blue) and
$y$-waves (black) for PhC structure from Fig. \ref{fig1}(a)
calculated with Berreman`s method. The parameters are $\epsilon_e
= 4$, $\epsilon_o = 1$, $d = 0.125 \ \mu m$, $(\Lambda - d) =
0.250 \ \mu m$, tilt angle $\phi = \pi/9$ (a), $\phi = \pi/18$ (b). 
}}
\label{fig1}
\end{figure*}

The system under scrutiny is composed of an ADL with two anisotropic PhC arms
attached to its sides as shown in Fig. \ref{fig1} (a). Each PhC
arm is a one-dimensional PhC with alternating layers of isotropic
and anisotropic dielectric materials. The layers are stacked along
the $z$-axis with period $\Lambda$. The isotropic layers are made
of a dielectric material with permittivity $\epsilon_o$ and
thickness $\Lambda-d$. The thickness of each anisotropic layer is
$d$. The anisotropic layers have their principal dielectric axes
aligned with the $x$, $y$-axes with the corresponding permittivity
component principal dielectric constants $\epsilon_{e}$,
$\epsilon_{o}$, but the principal axes of the ADL are tilted with
respect of the principal axes of the PhC arms as shown in Fig.
\ref{fig1}(a). Propagation of the monochromatic electromagnetic waves is
controlled by Maxwell's equations of the following form \cite{Timofeev18}

\begin{equation}
\left\{\begin{array}{cc}
0 & \nabla \times \\
-\nabla \times & 0
\end{array}\right\}
\left\{\begin{array}{c}
{\bm{E}} \\
{\bm{H}}
\end{array} \right\}=
-ik_0
\left\{\begin{array}{c}
\hat{\epsilon} {\bm{E}} \\
{\bm{H}} \end{array}\right\},
\label{Maxwell}
\end{equation}
where $\bm{E}$ is the electric vector, $\bm{H}$ is the magnetic vector, $k_0 = \omega/c$ is the wave number in vacuum
with $c$ as the speed of light, and, finally, $\hat{\epsilon}$ is the dielectric tensor.
The orientation of the ADL optical axis is
determined by the unit vector
\begin{equation}
\bm{a} = \{\cos{(\phi)}, \sin{(\phi)}, 0\}^{\dagger},
\label{a}
\end{equation}
as shown in Fig. \ref{fig1}(a).
Since the reference frame is aligned with the
optical axes in the PhC, the dielectric tensor is diagonal everywhere out of the ADL.
Given that $\bm{a}$ is specified by the tilt angle $\phi$, in the ADL it takes the following
form
\begin{equation}
\hat{\epsilon} = \left\{\begin{array} {cc}
\epsilon_e \cos^2 (\phi) + \epsilon_o \sin^2 (\phi) & \sin{(2\phi)} \;  (\epsilon_e-\epsilon_o)/2 \\
\sin{(2\phi)}  \; (\epsilon_e-\epsilon_o)/2 & \epsilon_e \sin^2 (\phi) + \epsilon_o \cos^2 (\phi)
\end{array} \right\}.
\label{eq:epsilon}
\end{equation}

In this paper we restrict ourselves with the normal incidence, i.e. the Bloch wave vector is
aligned with the $z$-axis in Fig. \ref{fig1}(a).  The dispersion of waves in the PhC arms depends on polarization.
For the $x$-polarized
waves ($x$-waves) the dispersion relationship is that of a one-dimensional PhC \cite{Rytov1956,
YarivYeh1984bk, ShiTsai1984_PBG, CamleyMills1984_PBG}
\begin{equation}
\cos{(K \Lambda)} = \cos{(k_{e} d)} \cos{[k_{o} (\Lambda - d)]} - \frac{1 + r_{o e}^2}{1 - r_{o e}^2} \sin{(k_{e} d)} \sin{[k_{o}
(\Lambda - d)]},
\label{Rytov}
\end{equation}
where $K$ is the Bloch wave number,
\begin{equation}
k_{e} = k_0 \sqrt{\epsilon_e} = k_0 n_e, \ k_{o} = k_0 \sqrt{\epsilon_o} = k_0 n_o,
\label{o_and_e}
\end{equation}
and the Fresnel coefficient $r_{o e}$ is given by
\begin{equation}
r_{o e} = \frac{k_{o} -  k_{e}} {k_{o} + k_{e}}.
\label{Fresnel_ab_perp}
\end{equation}
Equation (\ref{Rytov}) defines the band structure for the
$x$-waves with the condition  $|\cos{(K \Lambda)}| = 1$
corresponding to the edges of the photonic band gap in which
the wave propagation is forbidden. In Fig. \ref{fig1} (a) we
demonstrate the transmittance spectrum of the system with $20$
bi-layers in each PhC arms; the overall system being submersed in
air. One can see that for the $x$-waves the transmittance  is zero
within the band gap. On the other hand the PhC arms are always
transparent to the $y$-polarized waves ($y$-waves) with dispersion
$k_o=\epsilon_o k_0$. Notice, though, that the $y$-waves
transmittance exhibits a sharp dip at the center of the band gap. This
dip is due to a high quality resonant mode predicted in
\cite{Timofeev18}. Although the line shape is symmetric, the dip,
nonetheless, can be attributed to a Fano resonance as the
transmittance reaches zero at the center of the band gap indicating a
full destructive interference between two transmission paths.
In this paper we set a goal of finding the
analytic solution describing the Fano anomaly in the band gap.

\section{Resonant eigenmode and bound state in the continuum}

The resonant states are the eigenmodes of Maxwell's equations (\ref{Maxwell}) with reflectionless boundary conditions
in the PhC arms. The equation for resonant eigenfrequencies can be obtained by matching the general solution in the ADL
to the outgoing, both propagating and evanescent, waves in the PhC arms.
Let us start from the general solution in the ADL.

\subsection{General solution in the ADL}

First, the unit vector along the propagation direction is defined as 
\begin{equation}
\bm{\kappa}^{\scriptscriptstyle(\pm)}  = [0, 0, \pm 1],
\label{kappa_eo}
\end{equation}
where the symbol $\pm$ is used to discriminate between forward and backward propagating waves with respect to
the orientation of the $z$-axis. The ADL supports two types of electromagnetic
waves of different polarization. The $e$-waves with wavenumber $k_e=\epsilon_e k_0$ are polarized along the director $\bm{a}$,
equation (\ref{a}), while the
$o$-waves with wavenumber $k_o=\epsilon_o k_0$ have their electric vector perpendicular to both $\bm{a}$ and $\bm{\kappa}$,
as seen from Fig. \ref{fig1}.
The electric and magnetic vectors of the $e$-wave can be written as
\begin{equation}
\bm{E}_{e}^{\scriptscriptstyle(\pm)} = E_{e}^{\scriptscriptstyle(\pm)} \bm{a}, \ \ \bm{H}_{e}^{\scriptscriptstyle(\pm)}
= \frac{k_e}{k_0}\left[\bm{\kappa}^{\scriptscriptstyle(\pm)} \times \bm{E}_{e}^{\scriptscriptstyle(\pm)} \right],
\label{e_EH}
\end{equation}
where $E_{e}^{\scs (\pm)}$ are unknown amplitudes.
At the same time for $o$-waves we have
\begin{equation}
\bm{E}_{o}^{\scriptscriptstyle(\pm)} = E_{o}^{\scs (\pm)} \left[ \bm{a} \times \bm{\kappa}^{\scriptscriptstyle(\pm)} \right], \ \
\bm{H}_{o}^{\scriptscriptstyle(\pm)} = \frac{k_o}{k_0}\left[\bm{\kappa}^{\scriptscriptstyle(\pm)} \times \bm{E}_{o}^{\scriptscriptstyle(\pm)}
 \right],
\label{o_EH}
\end{equation}
where $E_{o}^{\scs (\pm)}$ are again unknown amplitudes.
The general solution of equations (\ref{Maxwell}) in the ADL, $\ z \in [-d,\ d]$, is written as a sum of
the forward and backward propagating waves
\begin{equation}
\bm{E} = \sum_{j=o,e} \left(\bm{E}_{j}^{\scriptscriptstyle(+)} e^{i k_{j} z} + \bm{E}_{j}^{\scriptscriptstyle(-)} e^{-i k_{j} z} \right),\ \
\bm{H} = \sum_{j=o,e} \left(\bm{H}_{j}^{\scriptscriptstyle(+)} e^{i k_{j} z} + \bm{H}_{j}^{\scriptscriptstyle(-)} e^{-i k_{j} z} \right).
\label{sum_EH}
\end{equation}

\subsection{General solution in the PhC arms}
The general solution of Maxwell's equations (\ref{Maxwell}) in the PhC arms is also written as a sum of forward
and backward  propagating waves. For
the $x$-waves the field components $E_x$ and $H_y$ in the isotropic layer
with the cell number $m$, $\ z \in [d + m \Lambda,\ (m + 1) \Lambda]$, are written as
\begin{equation}
\begin{aligned}
& E_{x}^{(m)} =  e^{iK \Lambda m}\left[A^{\scriptscriptstyle(+)}_x e^{i k_{o} (z - d - m  \Lambda)}
+ A^{\scriptscriptstyle (-)}_x e^{-i k_{o} (z - d - m \Lambda)}\right], \\
& H_{y}^{(m)} = \frac{k_{o}}{k_0} e^{iK \Lambda m}\left[A^{\scs (+)}_x e^{i k_{o}
 (z - d - m \Lambda)} - A^{\scs (-)}_x e^{-i k_{o} (z - d - m \Lambda)}\right].
\end{aligned}
\label{xEH1}
\end{equation}
In the anisotropic layer with the cell number $m$, $\ z \in [(m+1) \Lambda,\ d + (m + 1) \Lambda] $, we have
\begin{equation}
\begin{aligned}
& E_{x}^{(m)}  =  e^{iK \Lambda m}\left[B^{\scs (+)}_x e^{i k_{e} (z - m\Lambda-\Lambda)} + B^{\scs (-)}_x e^{-i k_{e} (z -  m\Lambda-\Lambda)}\right], \\
& H_{y}^{(m)} = \frac{k_{e}}{k_0} e^{iK \Lambda m}\left[B_x^{\scs (+)} e^{i k_{e} (z - m\Lambda-\Lambda)} - B_x^{\scs (-)} e^{-i k_{e} (z -  m\Lambda-\Lambda)}\right].
\end{aligned}
\label{xEH2}
\end{equation}
By applying the continuity condition for the tangential field components
the solutions (\ref{xEH1}) and (\ref{xEH2}) are matched
on the boundary between the anisotropic layer in the $(m-1)_{\rm th}$ cell and the isotropic
layer in $m_{\rm th}$ cell,  $\ z = d + m \Lambda$, as well as on the boundary between the layers
in the $m_{\rm th}$ cell, $\ z = (m + 1)\Lambda$. This gives us a system of four equations for four
unknowns, $A^{\scs (+)}, A^{\scs (-)}, B^{\scs (+)}, B^{\scs (-)}$. After solving for $B^{\scs (+)}$ and $B^{\scs (-)}$, this system can
be reduced to the following two equations
\begin{equation}
\left\{\begin{aligned}
A^{\scs (+)} \left[e^{i k_{o} (\Lambda - d)} - e^{iK \Lambda} e^{-i k_{e} d}\right] -
 A^{\scs (-)} r_{o e}\left[e^{-i k_{o} (\Lambda - d)} - e^{iK \Lambda} e^{-i k_{e} d}\right] = 0, \\
-A^{\scs (+)} r_{o e} \left[e^{i k_{o} (\Lambda - d)} - e^{iK \Lambda} e^{i k_{e} d}\right] +
A^{\scs (-)} \left[e^{-i k_{o} (\Lambda - d)} - e^{iK \Lambda} e^{i k_{e} d}\right] = 0,
\end{aligned}\right.
\label{PC_equations}
\end{equation}
where $r_{oe}$ is given by equation (\ref{Fresnel_ab_perp}). One can easily check that Eq. (\ref{PC_equations}) is only solvable
when $K$ satisfies the dispersion relationship (\ref{Rytov}).

In contrast to the $x$-waves, for the outgoing $y$-waves in the right PhC arms the solution is simple
\begin{equation}
\begin{aligned}
& E_{y}  =
-C^{\scs (+)} e^{i k_{o} (z - d)}, \\
& H_{x}  = \frac{k_{o}}{k_0} C^{\scs (+)} e^{i k_{o} (z - d)}.
\end{aligned}
\label{yEH}
\end{equation}
Notice that so far we have not written down the solution in the left PhC arm. The direct application of that solution can be avoided
by using the mirror symmetry of the system. Here, in accordance with ref \cite{Timofeev18} we restrict ourselves with the antisymmetric case
\begin{equation}
\bm{E}(z) = - \bm{E}(-z).
\label{atysymmetry}
\end{equation}
The generalization onto the symmetric case is straightforward.

\subsection{Wave matching}
Now we have all ingredients for finding the field profile of the antisymmetric resonant eigenmodes. By matching equation
(\ref{sum_EH}) to both equation (\ref{xEH1}) and equation (\ref{yEH}) on the interface between the ADL and the
right PhC arm and using equations (\ref{PC_equations}, \ref{atysymmetry}) one obtains eight equations for
eight unknown variables $E_{e}^{\scs(+)}, E_{e}^{\scs(-)}, E_{o}^{\scs (+)}, E_{o}^{\scs (-)}, A^{\scs (+)}, A^{\scs (-)}, C^{\scs (+)}, K$.
After some lengthy and tedious calculations one finds that the system is solvable under the following condition
\begin{equation}
\frac{\xi e^{i k_{o} (\Lambda - d)} - r_{o e} e^{-i k_{o} (\Lambda - d)}}{\xi e^{-i k_{e} d} - r_{o e} e^{-i k_{e} d}} -
\frac{e^{-i k_{o} (\Lambda - d)} - \xi r_{o e} e^{i k_{o} (\Lambda - d)}}{e^{i k_{e} d} - \xi r_{o e} e^{i k_{e} d}} = 0,
\label{disp}
\end{equation}
where
\begin{equation}
\xi = -e^{2i k_{o} d}\sin^2{(\phi)} + \frac{r_{oe} - e^{2i k_{e} d}}{1 - r_{oe} e^{2i k_{e} d}} \cos^2{(\phi)}.
\label{x}
\end{equation}
Taking into account equation (\ref{o_and_e}) we see that the above formulae represent the transcendental equation for
complex eigenvalues, $k_0$ of the Maxwell's equations (\ref{Maxwell}).
The analytic solution for the electromagnetic field within the ADL can be evaluated through the following formulae
\begin{equation}
\begin{aligned}
& E_{o}^{\scs (+)} = E_{o}^{\scs (-)}, \ E_{e}^{\scs (-)} = \zeta E_{o}^{\scs(-)}, \ E_{o}^{\scs (+)} = -\zeta E_{o}^{\scs (-)}, \ E_{o}^{\scs (-)} = \frac{i A} {2 n_e \zeta},\\
& \zeta = - \frac{e^{-i k_{o} d} t_{oe} \cos{(\phi)}}{(e^{-i k_{e} d} - r_{oe} e^{i k_{e} d}) \sin{(\phi)}}, \ t_{oe} = \frac{2k_{o}} {k_{o} + k_{e}}.
\end{aligned}
\label{fields}
\end{equation}
Substituting (\ref{fields}) into equation (\ref{sum_EH}) one finds the profile of the resonant eigenmode within the ADL
\begin{equation}
\begin{aligned}
& E_x = \frac{A}{n_e} \sin{(k_e z)} \cos{(\phi)} - \frac{A}{n_e \zeta} \sin{(k_o z)} \sin{(\phi)}, \\
& E_y = \frac{A}{n_e} \sin{(k_e z)} \sin{(\phi)} + \frac{A}{n_e \zeta} \sin{(k_o z)} \cos{(\phi)}.
\label{EADL}
\end{aligned}
\end{equation}
The amplitude $A$ has to be defined from a proper normalization
condition. We mention that in limiting case $\phi \rightarrow 0$
the $\zeta \rightarrow \infty$ and fields $E_x = (A/n_e) \sin{(k_e
z)}$, and $E_y = 0$ coincide with exact solution for BIC (8), (21)
from our previous work \cite{Timofeev18}. The obtained eigenfield
are plotted in Fig. (\ref{fig1}) (a). One can see that though the
$y$-component is localized due to the band gap, the $x$-component
grows with the distance from the ADL as it is typical for resonant
eigenstates \cite{Lalanne18}.

\subsection{Perturbative solution}
Equations (\ref{disp}, \ref{x}) are generally not solvable analytically. There is, however, a single tractable perturbative solution in
the case of quarter-wave optical thicknesses of the layers
\begin{equation}
k_{o} (\Lambda - d) = k_{e} d = \frac{k_0 \lambda_{\scs PBG}}{4} = \frac{\pi \omega}{2 \omega_{\scs PBG}},
\label{QW}
\end{equation}
where  $\omega_{\scs PBG}$ is the center frequency of photonic band gap, and $\lambda_{\scs PBG}$ is the corresponding wavelength.
In our previous work \cite{Timofeev18} we found an exact solution for $\phi=0$.
Here by applying a Taylor expansion of equations (\ref{disp}, \ref{x}) in powers
of the tilt angle $\phi$ we found approximate solutions for both resonant eigenfrequency and resonant eigenomode.
By writing the resonant eigenfrequency as
\begin{equation}
\omega_r=\tilde{\omega}-i \gamma,
\end{equation}
where both $\tilde{\omega}$ and $\gamma$ are real and positive,
and substituting into equations (\ref{disp}, \ref{x}) one finds
\begin{equation}
\begin{aligned}
& \tilde{\omega} = \omega_{\scs PBG} + \frac{\omega_{\scs PBG}}{\pi} q (1 - q) \sin{(\pi q)} \cdot \phi^{2} + {\cal O}(\phi^4)\\
& \gamma = \frac{2 \omega_{\scs PBG}}{\pi} q (1 - q) \cos^2{(\pi q/2)} \cdot \phi^{2}+ {\cal O}(\phi^4).
\label{omega}
\end{aligned}
\end{equation}
where $q = n_{o}/n_{e}$. Notice that the imaginary part of $\omega$ vanishes if $\phi=0$.
Thus, the system supports an antisymmetric BIC with the frequency $\omega_{\scs BIC}=\omega_{\scs PBG}$. That BIC
was first reported in our previous work \cite{Timofeev18}. We address the reader to the above reference for
detailed analysis of the BIC and the plots visualizing its eigenfields.
For the further convenience we also introduce the resonant vacuum wave number as
\begin{equation}
k_r = (\omega_r - i \gamma)/c = k_{\scs BIC}\left[1 + \alpha \cdot \phi^2+ {\cal O}(\phi^4)\right],
\label{k_BIC}
\end{equation}
where the complex valued $\alpha$ is implicitly defined by equation (\ref{omega}) and $k_{\scs BIC}=\omega_{\scs BIC}/c$.
Finally, expanding (\ref{EADL}) into the Taylor series in $\phi$ we find the following expression for the resonant eigenmode profile within the ADL
\begin{equation}
\bm{E}_r(z) = \frac{A}{n_e}
\left\{\begin{array} {c} \sin{\left(\frac{\pi z}{2 d}\right)}
+ {\cal O}(\phi^2)
 \\ \left[ \sin{\left(\frac{\pi z}{2 d}\right)} + ie^{\frac{i\pi q}{2}}  \sin{\left(\frac{\pi q z}{2 d}\right)} \right]\cdot \phi  + {\cal O}(\phi^3) \end{array}\right\}.
\label{E_ADL}
\end{equation}
Notice that ${\bm E}_r$ can be handled as $2\times 1$ vector since
$E_z=0$.

\subsection{Normalization condition}
There are several equivalent formulations of the FV normalization condition \cite{Muljarov10, Doost13, Muljarov18}.
Here we follow \cite{Doost13}, writing down the FV normalization condition through analytic continuation
${\bm E}(z,k)$ of the resonant eigenmode ${\bm E}_r(z)$ around the point $k=k_r$
\begin{equation}
\int\limits_{-d}^{d}{\bm E}_r\cdot\hat{\epsilon}{\bm E}_{r}dz
- \lim_{k\rightarrow k_r}\left\{ \frac{2\left[{\bm E}_r(d)\cdot\partial_z{\bm E}(d,k)-
{\bm E}(d,k)\cdot\partial_z{\bm E}_{r}(d) \right]}{k_{r}^2-k^2}\right\}=1.
\label{normalization}
\end{equation}
 At the first glance the "flux" term in equation (\ref{normalization})
differs from that in \cite{Doost13} by the factor of $2$; this is because the "flux" term is doubled to account for
both interfaces $z=\pm d$. For the resonant eigenmode $\bm{E}_r(z)$ found in the previous subsection the amplitude $A$ corresponding
equation (\ref{normalization}) can be found analytically
we the use of equation (\ref{E_ADL}). This would, however, result in a very complicated expression. Fortunately, we shall see later on that
in our case we do not need the general expression for $A$ in the second order perturabtive
solution consistent with equation (\ref{omega}).
By a close examination of equation (\ref{E_ADL}) one can see that the the Taylor expansion of $A$ can only
contain even powers of $\phi$.
Thus, for the further convenience we can write
\begin{equation}
A = \frac{1}{\sqrt{F+B\cdot\phi^2}}
\label{AB}
\end{equation}
assuming that $F$ and $B$ are such that the normalization condition (\ref{normalization}) is satisfied up to ${\cal O}(\phi^4)$.

\section{Scattering problem}

Let us assume that a monochromatic $y$-wave is injected into the system through the left PhC arm. The scattering problem can now be solved through the following
decomposition of the electric field within the ADL
\begin{equation}\label{decomposition}
{\bm E}={\bm E}_{\scs dir}+{\bm E}_{\scs res},
\end{equation}
where the direct contribution is simply the electric field of the incident wave
\begin{equation}\label{direct}
{\bm E}_{\scs dir}=\sqrt{I_0}e^{ik_oz}
\left\{0, \ 1\right\}^{\dagger}
\end{equation}
with intensity $I_0$, while ${\bm E}_{\scs res}$ can be viewed as the contribution due to the resonant pathway
via exitation of the resonant eigenmode ${\bm E}_r$. Substituting Eq. (\ref{decomposition}) into Eq. (\ref{Maxwell})
one obtains the inhomogeneous equation
\begin{equation}\label{inhomo}
\partial_z^2{\bm E}_{\scs res} +k^2_0\hat{\epsilon} {\bm E}_{\scs res}={\bm J}
\end{equation}
with
\begin{equation}\label{J}
{\bm J}=-\partial_z^2{\bm E}_{\scs dir} -k^2_0\hat{\epsilon} {\bm E}_{\scs dir}.
\end{equation}
Equation (\ref{inhomo}) can be solved with the use of Green's function
\begin{equation}\label{solved}
{\bm E}_{\scs res}(z)=\int\limits_{-d}^{d}\widehat{G}(z,z'){\bm J}(z')dz'
\end{equation}
that is defined as the solution of Maxwell's equations with a delta source
\begin{equation}\label{Green}
\partial_z^2\widehat{G}(z,z') +k^2_0\hat{\epsilon} \widehat{G}(z,z')=\delta(z-z')\widehat{\mathbb I}_{2},
\end{equation}
where $\widehat{\mathbb I}_{2}$ is the $2\times2$ identity matrix.  According to
\cite{Doost13} Green's function can expanded into the orthonormal resonant eigenmodes as
\begin{equation}\label{Green_expantion}
\widehat{G}(z,z')=\sum_n\frac{{\bm E}_n(z) \otimes{\bm E}_n(z')}{2k_0(k_0-k_n)}.
\end{equation}
Of course we do not know the full spectrum $k_n$, since equations
(\ref{disp}, \ref{x}) are not solved analytically. We, however,
assume that the contribution of all eigenfields except  ${\bm
E}_{\scs res}$ is accumulated in the direct field. Thus, in the
spectral vicinity of the quasi-BIC we apply the {\it resonant
approximation} taking into account only the eigenmode associated
with the BIC
\begin{equation}\label{Green_resonant}
\widehat{G}_{\scs res}(z,z')=\frac{{\bm E}_{r}(z) \otimes{\bm E}_{r}(z')}{2k_0(k_0-k_{r})}.
\end{equation}
The resonant field can now be calculated from equation (\ref{solved}) with
the resonant Green's function equation (\ref{Green_resonant}) once the FV normalization condition (\ref{AB})
is applied to the quasi-BIC eigenmode. The analytic expression for the resonant field reads
\begin{equation}
\bm{E}_{\scs res}(z) = \frac{1}{k_0(k_0-k_r) (F+B\cdot\phi^2) \epsilon_e}
\left[
\frac{i k_o \sqrt{I_0} k_0^2 (\epsilon_e-\epsilon_o) \cos{(k_o d)}}{k_{o}^2-\pi^2/4d^2} \cdot \phi
+ {\cal O}(\phi^3) \right]
\left\{\begin{array} {c} \sin{\left(\frac{\pi z}{2 d}\right)}
\\ \left[\sin{\left(\frac{\pi z}{2 d}\right)} + ie^{\frac{i\pi q}{2}} \sin{\left(\frac{\pi q z}{2 d}\right)} \right]\cdot \phi
   \end{array}\right\}.
\label{Solution}
\end{equation}
\begin{figure*}[h!]
\centering\includegraphics[scale=1]{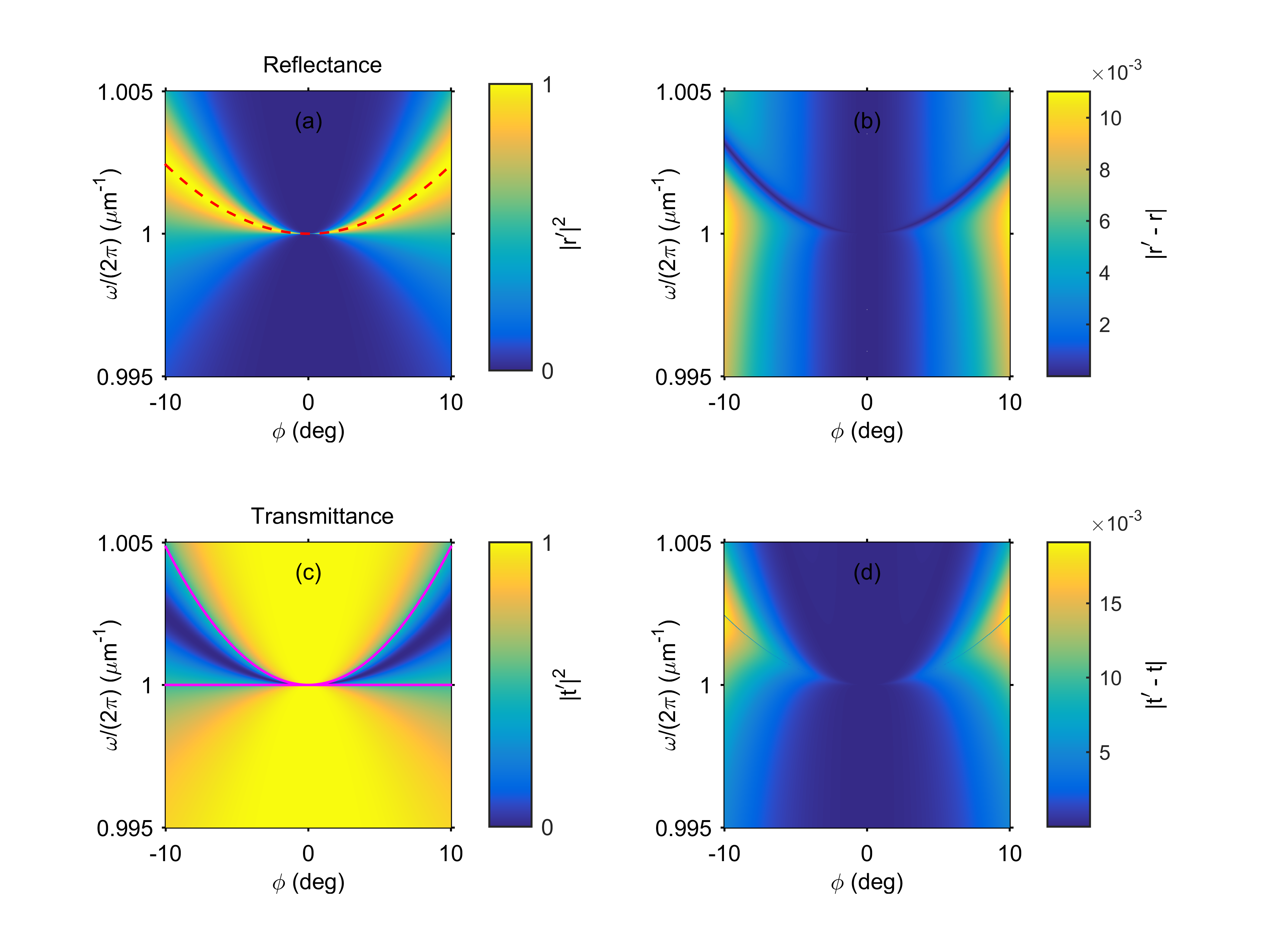} \caption{ {(a, c)
The reflectance  $|r^{\prime}|^2$ (a) and transmittance
$|t^{\prime}|^2$ (c) spectra vs tilt angle $\phi$ of $y$-waves for
PhC structure from Fig. \ref{fig1}(a) calculated with Berreman`s
method. The dashed red line shows the analytic resonant frequency
$\omega_r/(2\pi)$ \eqref{omega}. The solid magenta lines show the
analytic results for half-minima in transmittance $(\omega_r \pm
\gamma)/(2\pi)$.  (b, d) The difference between reflectance (b)
and transmittance (d) spectra calculated with Berreman`s method
and analytically \eqref{r}, \eqref{t}. The parameters are the same
as in Fig. \ref{fig1}.}} \label{fig2}
\end{figure*}
The above equation constitutes the perturbative solution of the
scattering problem with the accuracy up to ${\cal O} (\phi^3)$.
Notice that the terms dependant on $B$ also vanish as ${\cal O}
(\phi^3)$. Thus, in evaluating the FV normalization condition
(\ref{normalization}) we can restrict ourselves to $\phi=0$ in
which case the eigenmode is a BIC. Further on we safely set $B=0$
in all calculations. The BIC is localized function decaying with
$z\rightarrow \pm \infty$. Since the division into the scattering
domain and the waveguides is arbitrary and the flux term is
vanishing with $z\rightarrow \pm \infty$, one can rewrite the
normalization condition for the BIC as follows
\begin{equation}
\int\limits_{-\infty}^{\infty}{\bm E}_n\cdot\hat{\epsilon}{\bm E}_{n}dz=1.
\label{normalization_BIC}
\end{equation}
We mention in passing that the equivalence between equations (\ref{normalization}) and (\ref{normalization_BIC})
can also be proven by subsequently applying the Newton-Leibniz axiom and Maxwell's equations (\ref{Maxwell}) to the flux
term in equation (\ref{normalization}). The integral in equation (\ref{normalization_BIC}) is nothing but
the energy stored in the eigenmode up to a constant prefactor. This integral for the system in Fig. \ref{fig1}(a)
has been evaluated analytically in our previous work \cite{Timofeev18}. As a result the normalization condition (\ref{normalization_BIC})
yields
\begin{equation}
F=\frac{d}{1 - q}.
\label{A}
\end{equation}

Finally, the reflection and transmission coefficients can be found through the following
equations, correspondingly
\begin{equation}\label{r}
r=e^{-ik_od}{\bm e}_y^{\dagger} \cdot\left[{\bm E}(-d)-\sqrt{I_0}e^{-ik_od}{\bm e}_y\right],
\end{equation}
and
\begin{equation}\label{t}
t=e^{-ik_od}{\bm e}_y^{\dagger} \cdot {\bm E}(d),
\end{equation}
where ${\bm e}_y= \left\{0, 1 \right\}^{\dagger}$ and ${\bm E}$ is
computed through equations (\ref{decomposition}, \ref{direct},
\ref{Solution}). In Fig. \ref{fig2} we compare the perturbative
analytic solution against direct numerical simulations with the
Berreman transfer-matrix method \cite{Berreman72}. One can see
that deviation is no more than 2\% even for relatively large
angle $\phi = 10$ deg.

\section{Conclusion}

In this paper we considered light scattering by
an anisotropic defect layer embedded into anisotropic photonic crystal in the spectral vicinity of an optical BIC. Using a resonant state expansion method we derived an analytic solution for reflection and transmission amplitudes.
The analytic solution is constructed via a perturbative approach with the BIC as the zeroth order approximation.
The solution is found to accurately describe the collapsing Fano feature in the spectral vicinity of the BIC.
So far the theoretical attempts to describe the Fano feature induced by the BIC relied on phenomenological approaches such
as the $S$-matrix approach \cite{Blanchard16}, or the coupled mode theory \cite{Bulgakov18a}.
To the best
of our knowledge this is the first full-wave analytic solution involving an optical BIC reported in the literature.
We believe that the results presented offer a new angle onto the resonant state expansion method paving a way
to analytic treatment of resonant scattering due to optical BICs. In particular we expect that the resonant approximation
can be invoked to build a rigorous theory of nonlinear response \cite{Bulgakov19b}.
The BICs in photonic systems have already found important applications in
enhanced optical absorbtion \cite{Zhang15}, surface enhanced Raman spectroscopy \cite{Romano18}, lasing \cite{Kodigala17},
and sensors \cite{Romano18a}. We speculate that analytic results are of importance for a further insight into
localization of light as well as the concurrent phenomenon of collapsing Fano resonance.

\bibliography{BSC_light_trapping}



\section*{Acknowledgements}

This work was supported by Russian Foundation for Basic Research project No.~19-52-52006. This project is also supported by by the Higher
Education Sprout Project of the National Chiao Tung
University and Ministry of Education and the Ministry
of Science and Technology (MOST No. 107-2221-E-009-
046-MY3; No. 108-2923-E-009-003-MY3). 

\section*{Author contributions statement}

D.N.M. and T.I.V. conceived the idea presented. P.S.P. and D.N.M. have equally contributed to the analytic results.
P.S.P. ran numerical simulations. D.N.M., P.S.P., K.P.C. and T.I.V. have equally contributed to writing the paper.

\section*{Additional information}

\

\textbf{Competing interests}: The authors declare no competing interests.

\

\textbf{Data availability}:
The data that support the findings of this study are available from the corresponding author, P.S.P., upon reasonable request.

\end{document}

%% file: Fig_1a.pdf_tex
\begingroup%
  \makeatletter%
  \providecommand\color[2][]{%
    \errmessage{(Inkscape) Color is used for the text in Inkscape, but the package 'color.sty' is not loaded}%
    \renewcommand\color[2][]{}%
  }%
  \providecommand\transparent[1]{%
    \errmessage{(Inkscape) Transparency is used (non-zero) for the text in Inkscape, but the package 'transparent.sty' is not loaded}%
    \renewcommand\transparent[1]{}%
  }%
  \providecommand\rotatebox[2]{#2}%
  \newcommand*\fsize{\dimexpr\f@size pt\relax}%
  \newcommand*\lineheight[1]{\fontsize{\fsize}{#1\fsize}\selectfont}%
  \ifx\svgwidth\undefined%
    \setlength{\unitlength}{226.77165354bp}%
    \ifx\svgscale\undefined%
      \relax%
    \else%
      \setlength{\unitlength}{\unitlength * \real{\svgscale}}%
    \fi%
  \else%
    \setlength{\unitlength}{\svgwidth}%
  \fi%
  \global\let\svgwidth\undefined%
  \global\let\svgscale\undefined%
  \makeatother%
  \begin{picture}(1,0.75)%
    \lineheight{1}%
    \setlength\tabcolsep{0pt}%
    \put(0,0){\includegraphics[width=\unitlength,page=1]{Fig_1a.pdf}}%
    \put(0.83999805,0.20936279){\color[rgb]{0,0,0}\makebox(0,0)[lt]{\lineheight{1.25}\smash{\begin{tabular}[t]{l}$x$\end{tabular}}}}%
    \put(0.88424911,0.08443489){\color[rgb]{0,0,0}\makebox(0,0)[lt]{\lineheight{1.25}\smash{\begin{tabular}[t]{l}$y$\end{tabular}}}}%
    \put(0.86675215,0.17949448){\color[rgb]{0,0,0}\makebox(0,0)[lt]{\lineheight{1.25}\smash{\begin{tabular}[t]{l}$z$\end{tabular}}}}%
    \put(0.12977343,0.50911446){\color[rgb]{0,0,0}\makebox(0,0)[lt]{\lineheight{1.25}\smash{\begin{tabular}[t]{l}$\epsilon_e$\end{tabular}}}}%
    \put(0.22240243,0.41183357){\color[rgb]{0,0,0}\makebox(0,0)[lt]{\lineheight{1.25}\smash{\begin{tabular}[t]{l}$\epsilon_o$\end{tabular}}}}%
    \put(0.37564731,0.49769859){\color[rgb]{0,0,0}\makebox(0,0)[lt]{\lineheight{1.25}\smash{\begin{tabular}[t]{l}$\epsilon_o$\end{tabular}}}}%
    \put(0.30235538,0.56842346){\color[rgb]{0,0,0}\makebox(0,0)[lt]{\lineheight{1.25}\smash{\begin{tabular}[t]{l}$\epsilon_o$\end{tabular}}}}%
    \put(0.55428157,0.64793606){\color[rgb]{0,0,0}\makebox(0,0)[lt]{\lineheight{1.25}\smash{\begin{tabular}[t]{l}$\bm{a}$\end{tabular}}}}%
    \put(0.49099951,0.58713989){\color[rgb]{0,0,0}\makebox(0,0)[lt]{\lineheight{1.25}\smash{\begin{tabular}[t]{l}$\phi$\end{tabular}}}}%
    \put(0,0){\includegraphics[width=\unitlength,page=2]{Fig_1a.pdf}}%
    \put(0.09522272,0.5586956){\color[rgb]{0,0,0}\makebox(0,0)[lt]{\lineheight{1.25}\smash{\begin{tabular}[t]{l}$d$\end{tabular}}}}%
    \put(0,0){\includegraphics[width=\unitlength,page=3]{Fig_1a.pdf}}%
    \put(0.14948503,0.64062042){\color[rgb]{0,0,0}\makebox(0,0)[lt]{\lineheight{1.25}\smash{\begin{tabular}[t]{l}$(\Lambda - d)$\end{tabular}}}}%
    \put(0,0){\includegraphics[width=\unitlength,page=4]{Fig_1a.pdf}}%
    \put(0.41300159,0.71033878){\color[rgb]{0,0,0}\makebox(0,0)[lt]{\lineheight{1.25}\smash{\begin{tabular}[t]{l}$2d$\end{tabular}}}}%
  \end{picture}%
\endgroup%